\documentclass[11pt,twoside]{article}

\usepackage{asp2006}
\usepackage{epsf}
\usepackage{graphicx}
\usepackage{sidecap}

\def\ion#1#2{{\rm #1}\,{\footnotesize\sc{#2}}\relax}

\begin{document}
\title{Photoionization Models of the Broad-line Region}   
\author{Karen M.\ Leighly\altaffilmark{1}  and Darrin
  Casebeer}   
\affil{Homer L.\ Dodge Department of Physics and Astronomy, The  
  University of Oklahoma, 440 W.\ Brooks St., Norman, OK 73019}    
\altaffiltext{1}{Visiting Professor, The Ohio State University,
  Department of Astronomy, 4055 McPherson Laboratory, 140 West 18th
  Avenue, Columbus, OH 43210-1173}

\begin{abstract} 

The strong broad emission lines in the optical and UV spectra of
Active Galactic Nuclei (AGN) are important for several reasons.  Not
only do they give us information about the structure of the AGN, their
properties are now used to estimate black hole masses and
metallicities in the vicinity of quasars, and these estimates are
propagated widely throughout astronomy today.  Photoionization codes
such as {\it Cloudy} are invaluable for understanding the physical
conditions of the gas emitting AGN broad lines.   In this review, we
discuss briefly the development of the historical 
``standard'' model.  We then review evidence that the  following  are
important: 1.) the column density, in particular the presence of gas
optically thin to the hydrogen continuum, influences the line
emission; 2.) the BLR emission  region is comprised of gas with a
range of densities and ionization parameters; 3.) the spectral energy
distribution of an individual AGN influences the line ratios in an
observable way.  
\end{abstract}


\section{Introduction and Motivation}

Strong, broad emission lines are an identifying feature of Active
Galactic Nuclei (AGN), and the line emission has been very important
in understanding their nature from the beginning.  A classic example
is the discovery that the emission lines observed in the spectrum of
the star-like radio object 3C~273 were Balmer lines with a large
redshift, which led to the realization that quasars
are very luminous, cosmological objects.  The study of the broad
emission lines in AGN and quasars remains relevant today.  The lines
give us information about the central engine, as they are 
broadened through bulk motions, either virial or acceleration or
both. The resulting emission-line widths are used to infer the black
hole mass.  These black hole mass estimates are now being used
throughout astronomy; for example, they are being used to study the
coevolution of the black holes and host galaxies.  Second, the line
properties are important components of Eigenvector 1 (E1), the  strong
set of correlations first observed by \citet{bg92} and then extended
by a number of workers \citep[e.g.,][]{wills99}. Eigenvector 1
demonstrates that AGN properties are correlated on a large range of
scales, from the X-ray emission at tens of Schwarzschild radii, to the
broad line region on 0.1 parsec scales, and the narrow-line region,
and possibly beyond. It is currently thought that E1 reflects the
accretion rate relative to the Eddington value.  Third, the line
emission offers a probe of the metallicity of the line-emitting gas,
which in turn may provide a measure of the chemical evolution in the
galaxy hosting the quasar, and the chemical evolution of the
Universe. Thus we see that the study of AGN broad lines is relevant in
astrophysics today.

Photoionziation modeling of the broad-line region is also a mature
discipline, but is likewise still very important.  It is easy to see
that it is even if we limit ourselves to only the uncertainties in the
applications in emission-line studies mentioned above.  For example,
the number of objects that have been adequately reverberation-mapped
is small; only four objects have \ion{C}{IV} lags.  Yet the results
are applied almost indiscriminately throughout astrophysics, even to
objects that have different properties (e.g., much lower or higher
luminosities or accretion rates).  Second, although we believe that E1
is a consequence of the accretion rate, it is important to realize
that there is currently no detailed model that can predict
emission-line properties given the accretion rate and black hole
mass. Finally, using line  ratios as a metallicity probe depends
directly on whether the ratios are robust in the face of a range of
gas conditions. Photoionization modeling can be used to address all of
these problems. 
 
We present here a review of photoionization modeling of the broad-line
region, but note this vast topic cannot be thoroughly explored in the
limited space available.  We first discuss the observations and
modeling that lead to the ``standard model'', and important
developments after that.  We then discuss three relatively recent
developments in observations and photoionization modeling of AGN
emission lines: optically thin gas, the LOC model, and the effect of
the spectral energy distribution.

\section{The Early Years and Development of the Standard Model}

Photoionization models for quasar emission lines were constructed
rapidly after their discovery because, at that time, photoionization
models for nebulae in our Galaxy were already well developed.  Early
models were quite limited as a consequence of the limitations in the
observations.  For example, there was no strong evidence for
differences in profiles of emission lines, so it was assumed that the
lines were emitted by the same gas, and that gas was modelled using a
single ionization parameter and density (hereafter referred to as the
one-zone model).

In the one-zone model, the observed line emission depends principally
on the ionization parameter $U=\Phi/nc$, which defines the balance
between the photoionization rate determined by the photon flux $\Phi$
and the recombination rate determined by the density $n$, normalized
by the speed of light $c$.  $U$ was constrained by line ratios;
generally the strengths of \ion{C}{IV} and \ion{C}{III}] relative to
Ly$\alpha$ were used. Ionization parameters were estimated to be $-2.8
< \log(U) < -1.5$.  The lower limit on the density was obtained from
the lack of a broad component of [\ion{O}{III}], and the upper limit
was constrained by the presence of \ion{C}{III}] to be less than about
$10^{10}\rm\, cm^{-3}$.  The covering fraction of the line emitting
gas was inferred from the Ly$\alpha$ equivalent width and the slope of
the ionizing continuum by counting the number of ionizing photons and
comparing with the observed Ly$\alpha$; it was found to be about 10\%
\citep[e.g.,][]{bn78}.  The shape of the ionizing continuum was based
on extrapolation of observed continua, but in addition, the continuum
could also be constrained from the ratio of \ion{He}{II} lines and the
hydrogen lines, as that ratio depends on the relative flux in the
\ion{He}{II} continuum (i.e., $>54.4\rm \, eV$) versus the Lyman
continuum \citep{williams71}.

Initially, models could not explain the relatively small ratio of
Ly$\alpha$ to H$\beta$. It was then discovered that the relatively
hard spectral energy distribution characteristic of AGN produced a
partially-ionized zone beyond the hydrogen ionization front.   In this
region, Ly$\alpha$ is optically thick, but the opacity to the Balmer
emission is lower, so the predicted Ly$\alpha$/H$\beta$ ratio was
smaller.  Assuming that the ratio of the weak \ion{C}{II}]~$\lambda
2325$ line to Ly$\alpha$ is 1/30, the column density of the emitting
region was constrained to be $10^{23}\rm\, cm^{-3}$ \citep{kk81}.   

Thus, the ``Standard Model'' was defined by the following
parameters: ionization parameter $\log U \sim -1.5$, density
$n \sim 4 \times 10^{9}\rm \,cm^{-3}$, covering fraction $\sim
0.1$, and column density $N_H \sim 10^{23} \rm \, cm^{-3}$
\citep{kk81}.    

A major modification of the standard model came in the mid-1980's when
the results of the first large-scale reverberation mapping campaigns
were reported. The results were surprising as the emission lines
responded much more quickly to changes in the continuum than
anticipated.  This has been interpreted as evidence that line
emission occurs much closer to the central engine than inferred from
the standard model, and therefore to maintain the same line ratios
(controlled primarily by $U$), the BLR gas must be much denser than
previously thought.    

The problem of photoionization at high densities was considered by
\citet{rnf89}. The physics of the gas is different at high densities;
as noted by \citet{rnf89}, they were approaching stellar atmosphere
conditions from the nebular limit.  The emission from gas at very high
densities doesn't look very much like that observed from AGN.   At
high densities, the familiar strong UV lines such as 
Ly$\alpha$, \ion{C}{IV} and especially \ion{C}{III}] become
thermalized. Continuum emission (Balmer, Paschen and free-free)
dominates at the highest densities; these continua are generally not
observed to be strong. Initially, this was thought to be a problem for
photoionization models.  However, \citet{ferland92} noted that in
NGC~5548, the density-sensitive line \ion{C}{III}] lagged the
continuum by a much longer time (3--4 weeks) compared with
\ion{C}{IV} and Ly$\alpha$ ($\sim 8 \rm \, days$), which have much
higher critical densities.  Thus, these authors were able to construct
a model with $\log n \sim 10^{11}\rm \, cm^{-3}$ and $\log(U)\sim -1$.

At the same time that the first reverberation-mapping results were
coming out, S.\ Collin and co-workers were working on a stratification
of the broad-line region of another kind \citep[for a summary,
see][]{cl88}.  They pointed out that simultaneously producing high-
and low-ionization emission in the same cloud was difficult;
specifically, under physical conditions that produce adequate
high-ionization lines such as \ion{C}{IV}, the amount of
low-ionization lines such as \ion{Mg}{II}, \ion{C}{II}, \ion{Fe}{II} and
the Balmer and Paschen continua would be underestimated, and vice
versa.  That can be reconciled if there are two different emitting
regions which see different continua.  

Observational support for two emission regions comes from evidence
that the high-ionization lines tend to be broader and blueshifted with
respect to the low-ionization lines.  One of the most
spectacular examples of this kinematic separation is seen in the two
NLS1s IRAS~13224$-$3809 and 1H~0707$-$495 \citep{lm04}.
In these objects, both the low-ionization lines 
(\ion{Mg}{II} and the Balmer lines) and the intermediate ionization
lines (\ion{C}{III}], \ion{Si}{III}] and \ion{Al}{III}) are narrow and
symmetric, while the high-ionization lines, including \ion{C}{IV}
and \ion{N}{V} are broad and blueshifted. Note that this trend is not
universal; in some objects the emission lines have almost consistent
profiles independent of ionization potential \citep[e.g.,
RE~1034$+$39;][]{clb06}.  In addition, \citet{bl05} report that
\ion{C}{IV} is narrower than the H$\beta$ in objects with broad
H$\beta$. 

\section{New Developments}

In recent years there have been
several significant advances in broad-line region modeling.
Unfortunately, we do not have space to discuss them all.  The most
important one that we do not discuss is the effects of
microturbulence, which changes line ratios by changing the opacity, in
essence \citep[e.g.,][]{bottorff02}.  Also important is the effect of
metallicity which changes the line emission by changing the cooling
rates \citep[e.g.,][]{sg99}.

\subsection{Optically-thin Broad-line Region Gas}

The column density in the standard model is  $10^{23}\rm\, cm^{-2}$
for an ionization parameter of 0.03.  These parameters imply that the
emitting region is ionization bounded; i.e., the hydrogen ionization
front is present in the emitting region.   

There is plenty of observational evidence for the presence also of gas
that is matter-bounded and thus optically thin to the hydrogen continuum
\citep{shields95}.   For example, in Fairall~9, the \ion{C}{IV}
luminosity saturates when the object is in a high state; this is
explained if optically-thin  \ion{C}{IV} emitting clouds become
completely ionized at high luminosities.  

Other evidence comes from profile studies.  For example,
\citet{zheng92} compared the profiles of H$\alpha$ and Ly$\alpha$, and
inferred that the larger width observed in Ly$\alpha$ could originate
in an inner region of optically-thin gas.  Both \citet{ferland96} and
\citet{leighly04} infer that the outflowing gas responsible for
blueshifted emission lines is optically thin.  They base their
inference on the fact that the high-ionization lines are blueshifted,
but the low-ionization lines aren't.  It makes sense that outflowing
gas be optically thin to the Lyman continuum if it is accelerated by
radiation pressure.  As discussed by \citet{al94}, the continuum is
attenuated as it penetrates deeper into the gas, and thus its ability
to accelerate the ions via resonance scattering decreases, especially
when the hydrogen ionization front is crossed.

Since different lines form at different depths in a photoionized slab
\citep[e.g., Fig.\ 2 in][]{hamann02}, the observed emission-line
fluxes in matter-bounded gas are a strong function of column density.
Depending on how deep one integrates into the slab can change, for
example, the \ion{N}{V}/\ion{C}{IV} ratio, since \ion{C}{IV} has a lower
ionization potential and is produced at larger depths than \ion{N}{V}.  

\subsection{The LOC Model}

The locally-optimally emitting cloud (LOC) model was first proposed by
\citet{baldwin95}.  This model is based on a very simple and
attractive premise.  The ``standard'' BLR model proposes that BLR
emission lines are best modelled by gas with particular values of the
ionization parameter and density.  Realistically, though, the emission
is more likely to be the average of gas with a range of ionization
parameters and densities. \citet{baldwin95} pointed out that any given
line is emitted most intensely (optimally) by gas with a particular
value of ionization parameter and density.  Thus, a weighted average
over density and ionizing flux includes gas with the optimal
parameters for a range of lines depending on their ionization
potential and critical densities. It is worth pointing out that
\citet{baldwin95} were not the first to average emission from gas with
a range of properties; their insightful contribution was that because
of the optimal-emitting properties, the predicted emission should not
be very dependent on the averaging procedure.

The LOC model allows us to discard the concept of fitting spectra by
locating the best ionization parameter and density. Nevertheless,
there are a number of parameters that need to be constrained, as
follows.  The spatial (radial) and density distributions must be
determined; assuming that these are power laws, the indices must be
specified.  In addition, the inner and outer radii, the lowest and
highest densities, the column density of the gas, and the illuminating
spectral energy distribution must all be specified. In addition, if
line equivalent widths are computed, a global covering fraction must
also be specified. 

A nominal parameter set has been used by \citet{hamann02} and others: 
$\gamma =-1$ and $\beta= -1$, where $\gamma$ and $\beta$ 
are the power law indices for the radial and density distributions,
respectively, $17 < \log \Phi < 24$ and $7 < \log n < 14$, where $\Phi$ and
$n$ are the photoionizing flux and hydrogen density, and
$N_{H}=10^{23}\rm \, cm^{-3}$.  In Fig.~1,
we explore the consequences of deviating from these standard
parameters.  The data and errors shown are the means and standard
deviations of strong emission lines or features measured from
composite spectra constructed by \citet{francis91, zheng97,
  brotherton01, vandenberk01}.  The global covering fractions are
obtained by fitting the LOC model results to the means and standard
deviations.  We find that the nominal parameters fit relatively well,
although the simulated \ion{Mg}{II} is too strong.  We also find that 
changing the SED can change the LOC prediction significantly, which
shows that SED effects that we will discuss in \S 3.3 are robust in the
face of averaging \citep[see   also][]{clb06}. 

\begin{figure}[t]
\begin{center}
\includegraphics[width=4.5in]{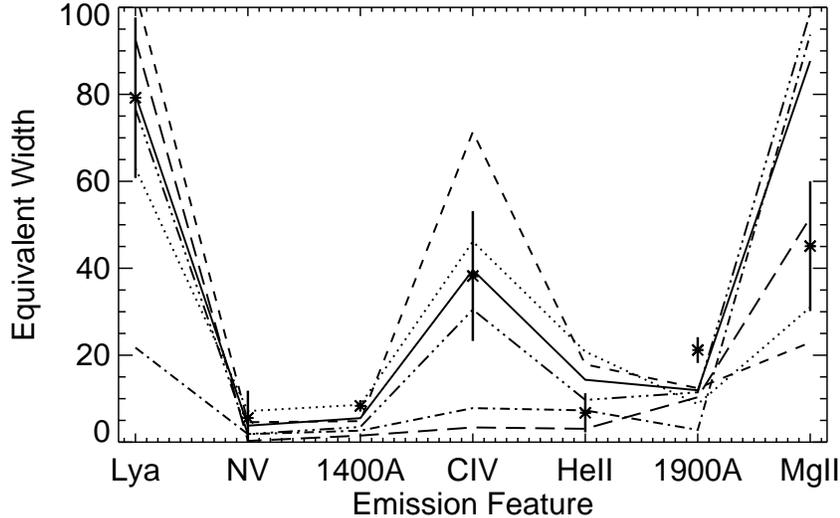}
\vskip -0.5pc
\caption{\small 
{An illustration of dependence of LOC on parameters.  Points with
  error bars are means and standard deviations of equivalent-width
  measurements from composite spectra.  Lines show the LOC models,
  scaled to the best fitting covering fraction.  Solid: nominal
  model ($C_f=0.24$); dotted: smaller outer radius ($C_f=0.46$); short
  dash: weighted toward low density ($C_f=0.46$); dash-dot: weight
  toward high density ($C_f=0.46$); dash-dot-dot: weighted toward small
  radii (0.87); long dashes: PHL 1811 SED ($C_f=1.0$).}}
\label{fig1}
\end{center}
\end{figure}

Despite the success of the model, it should be noted that the
fitted parameters cannot be interpreted physically; i.e., they don't
yield information about the structure and kinematics of the gas
emitting the broad lines.  In addition, there are a number of physical
effects that are not included.  A significant one is that the model
does not consider self-shielding; i.e., each cloud sees 
the continuum directly.  As discussed in \S 3.3., self-shielding (what
we call ``filtering'') may be important.  Also, in principle, many of
the parameters could also depend on radius, including the global
covering fraction, the column density, and the density-distribution
index, etc.

The LOC model is now used widely.  For example, \citet{kg00}
wanted to see if an LOC model constructed to fit the mean {\it HST}
spectrum from the NGC~5548 monitoring campaign would also be able
to explain the emission-line lags.  The LOC model was constructed by
fitting the outer radius, the radial distribution power law index, and
the global covering fraction.  The light curves were obtained by 
folding the UV continuum light curve with the emission-line response
functions constructed from the resulting LOC model. The agreement
between the observed and simulated lags for \ion{C}{IV} and Ly$\alpha$
was quite good, although simulated lags were longer for \ion{He}{II}
and \ion{N}{V},  and shorter for \ion{C}{III}] and \ion{Mg}{II} than
  the observed lags. There were also some differences in amplitude of
  the variations. Nevertheless, the agreement is impressive and
  provides important support for the model.

The LOC model has also been used extensively in quasar abundance
studies.  \citet{hamann02} used an LOC to determine which line ratios
were the most robust abundance indicators.  \citet{nmm06} constructed 
quasar composite spectra as a function of luminosity and redshift, and
interpreted the line ratios using an LOC model in which they optimized
the density and radial distribution power law indices.  An LOC model
has recently also been used to interpret the spectra of two
nitrogen-rich quasars found in the SDSS \citep{dhanda06}.  They find
that while one of their quasars is well described using the standard
LOC, the other has unusual low-ionization lines similar to the
prototypical NLS1 I~Zw~1, and the standard LOC model yielded less
consistent results among various metallicity indicators.
\citet{dhanda06} speculate that non-standard LOC parameters may be the
cause of the discrepancy.  Alternatively, this may be an example where
self-shielding is important. \citet{leighly04} analyzed {\it HST}
spectra from two NLS1s rather similar to I~Zw~1 (IRAS~13224$-$3809 and
1H~0707$-$495) and found that the intermediate- and low-ionization
line emission could be more consistently modelled if the continuum
illuminating it had first transversed the wind that emits the
blue-shifted high-ionization lines. This ``filtering'' concept will be
discussed further in \S 3.3. 

\subsection{The Spectral Energy Distribution}

There are a range of spectral energy distributions (SEDs) observed
among AGN. Perhaps the clearest evidence for this is the persistent
observation that $\alpha_{ox}$\footnote{$\alpha_{ox}$ is the
point-to-point slope   between 2500\AA\/ and $2\rm \, keV$ in the rest
frame.} is correlated with the UV luminosity such that more luminous
objects have steeper $\alpha_{ox}$'s \citep[e.g.,][]{steffen06}.  The
spectral energy distribution should be a function of a number of
parameters, including the black hole mass, the 
accretion rate, the inclination of the accretion disk, and the
fraction of energy channeled to the disk versus the corona.  Explicit
links between these parameters and observed SEDs have been 
elusive, partially because the extreme-UV portion of SED cannot
generally be observed.

The spectral energy distribution should influence the emission lines
from AGN, simply because the ions involved have a range of ionization
potentials.  But the influence of the SED is expected to be a
secondary effect because the lines emitted by photoionized gas are
controlled primarily by the ionization parameter
\citep[e.g.,][]{kk88}.   One of the most compelling pieces of evidence
that the SED is important is the observation that the slope of the
Baldwin effect is steeper for high-ionization lines than for low
ionization lines \citep[e.g.,][]{dietrich02}.  This is interpreted
appealingly as evidence that, as the luminosity increases, the
spectral energy distribution softens, resulting in a more rapid
decrease in the highest ionization lines.

In recent years, we have focused on observations and modelling of the
emission lines of Narrow-line Seyfert 1  galaxies (NLS1s), and we have
found a number of interesting links between the SEDs and the observed
emission lines. 

\subsubsection{RE~1034$+$39 -- A Very Hard SED}

RE~1034$+$39 is a Narrow-line Seyfert 1 galaxy that is well known
for its unusual spectral energy distribution.  While the optical and
UV spectra of the typical AGN rises toward the blue, forming the big
blue bump, RE~1034$+$39's optical/UV spectrum is rather red.  The big
blue bump appears instead in the X-ray spectrum: it appears to peak at
$\sim 100 \rm \, keV$, and the high energy turnover toward long
wavelengths appears to be in the soft X-rays \citep{pmsp95}.

In 1999, we performed coordinated observations of RE~1034$+$39 using
{\it FUSE}, {\it EUVE}, and {\it ASCA}.  We then analyzed the line
emission from the {\it FUSE} observation and an archival {\it HST}
observation \citep[please see][for details]{clb06}.  The emission
lines were modelled well by single Lorentzian profiles centered at
the rest wavelengths; no significant blueshift was observed in the
high-ionization lines, including \ion{O}{VI}.  This meant that we could
assume, at least to first approximation, that all the lines were
produced by the same or similar gas.   

A strong \ion{O}{VI} line was present in the {\it FUSE} spectrum. 
O$^{+5}$ has an ionization potential in the soft X-rays
(113~eV). Thus, we can test whether the emission lines are influenced
by the unusually hard SED.
We modelled the emission lines in this object in several different
ways.  We first used a single-zone model and confirmed that the line
emission is consistent with the hard SED.  Next, we computed an LOC
model using the observed SED, and we compared the results with those
using a standard SED.   The LOC model using the RE~1034$+$39 SED
produced enhanced \ion{O}{VI} as observed, but also yielded far too
strong \ion{Mg}{II}.   Finally,  we develop a series of semi-empirical
SEDs, ran {\it Cloudy}  models, and compared the results with the
measured values using a figure of merit (FOM).  The FOM minimum
indicates similar SED and gas properties as were inferred from the
one-zone model using the RE~1034$+$39 continuum. More importantly, the
FOM increases sharply toward softer continua, indicating that a hard
SED is {\it required} by the data in the context of a one-zone model.

\subsubsection{PHL~1811 -- A Very Soft SED}

PHL~1811 is a nearby ($z=0.192$), luminous
($M_B=-25.5$) narrow-line quasar \citep{leighly01}.  It is extremely
bright (B=14.4, R=14.1); it is the second brightest quasar at $z>0.1$
after 3C~273.  However, it was not detected in the ROSAT All Sky
Survey.  A pointed {\it BeppoSAX} observation  detected the object,
but yielded too few photons to unambiguously determine the cause of
the X-ray weakness; \citet{leighly01} speculated that either it is
intrinsically X-ray weak, or it is a nearby broad-absorption line
quasar and the X-ray emission is absorbed, or it is highly variable,
and we caught it both times in a low state.  

In 2001, we obtained coordinated {\it HST} and two {\it Chandra}
observations of PHL~1811 \citep{leighly07a, leighly07b}. The two {\it
Chandra} observations and an {\it XMM-Newton} observation obtained in
2004 reveal a weak X-ray source with a steep spectrum and no evidence
for intrinsic absorption.   Variability by a factor of four between
the two {\it Chandra} observations separated by 12 days 
suggest that the X-rays are not scattered emission.  No intrinsic
absorption lines were observed in the {\it HST} UV spectra, so we can
firmly rule out the idea that PHL~1811 is a BALQSO.  The X-ray spectra
are simple power laws with photon indices $\Gamma=2.3 \pm 0.1$; it
appears that we observe the central engine X-rays directly with no
evidence for reprocessing.  Including two recent {\it Swift} ToO
snapshots, PHL~1811 has been observed in X-rays a total of seven times
since 1990, and it has always been significantly X-ray weak with
$\alpha_{ox}=-2.3\pm 0.1$.  While we can never rule out the idea that
we coincidentally always observe it in a low state, the probability
that this is true is decreasing.  We conclude that that PHL~1811 is
intrinsically X-ray weak \citep{leighly07a}.   

The optical and UV spectra of PHL~1811 are very interesting
\citep{leighly07b}.  The continuum is as blue as an average quasar.
There is no evidence for forbidden or semiforbidden lines.  The
near-UV spectrum is dominated by \ion{Fe}{II} and \ion{Fe}{III}, and
unusual low-ionization lines such as \ion{Na}{I}~D and
\ion{Ca}{II}~H\&K are present. The high-ionization lines are very
weak; \ion{C}{IV} has an equivalent width of 6.6\AA\/, a factor of
$\sim 5$ smaller than that measured in quasar composite spectra. An
unusual feature near 1200\AA\/ can be deblended in terms of
Ly$\alpha$, \ion{N}{V}, \ion{Si}{II} and \ion{C}{III}* using the
blueshifted \ion{C}{IV} profile as a template.   

It turns out that the unusual line emission can be explained well by
the X-ray-weak SED.  The soft SED lacks high energy photons, and
therefore high-ionization line emission should be weak, as observed.
The average energy of a photoionizing photon is 
low, resulting in a low electron temperature, which means that
collisionally-excited lines should be weak.  Lines effected by
the cool temperature include permitted lines like \ion{Mg}{II}, but
also semiforbidden lines such as \ion{C}{III}] and \ion{Si}{III}].  These
semiforbidden lines have critical densities typical of gas densities
in AGN, and they are generally used as density diagnostics.
Thus, when the SED is soft, the semiforbidden lines fail as density
diagnostics. The gas ends up being ``cooling-challenged'', and has
properties similar to a high-density gas, such as strong continua due
to the inability of the gas to cool by producing the usual
collisionally-excited line emission.  Despite the low temperature, the
H$^+$ fraction in the partially-ionized zone is relatively high.  The
number of hydrogen in $n=2$ is relatively high because of the strong
diffuse emission, and these are ionized by the Balmer continuum, which
is relatively brighter for a soft SED compared with a hard SED for the
same ionizing photon flux.  The relatively strong UV continuum means
that lines that are pumped by the continuum, (e.g., \ion{Fe}{II} and
high-excitation \ion{Si}{II}) are stronger.  Thus conditions in the
gas are much different when the SED is very soft \citep{leighly07b}.

\subsubsection{The Filtered Continuum}

One of the biggest challenges in understanding the influence of the
SED is that the line-emitting gas may not see the same continuum that
we see.  On the other hand, if we can use the lines to estimate the
SED, we may obtain information about the geometry of the central
engine and line-emitting gas.  One way that the SED that the
line-emitting gas sees is different than the SED we see is if the
continuum is transmitted through ionized gas before it illuminates the
line-emitting gas.  This is a process that we call ``filtering'' and
note that it should be differentiated from ``shielding'' gas as
discussed by \citet{mcgv95} in which the continuum is attenuated but
the shape not explicitly changed.  We illustrate the concept of
filtering in Fig.\ 2.


\begin{figure}[t]
\begin{center}
\includegraphics[width=3.5in]{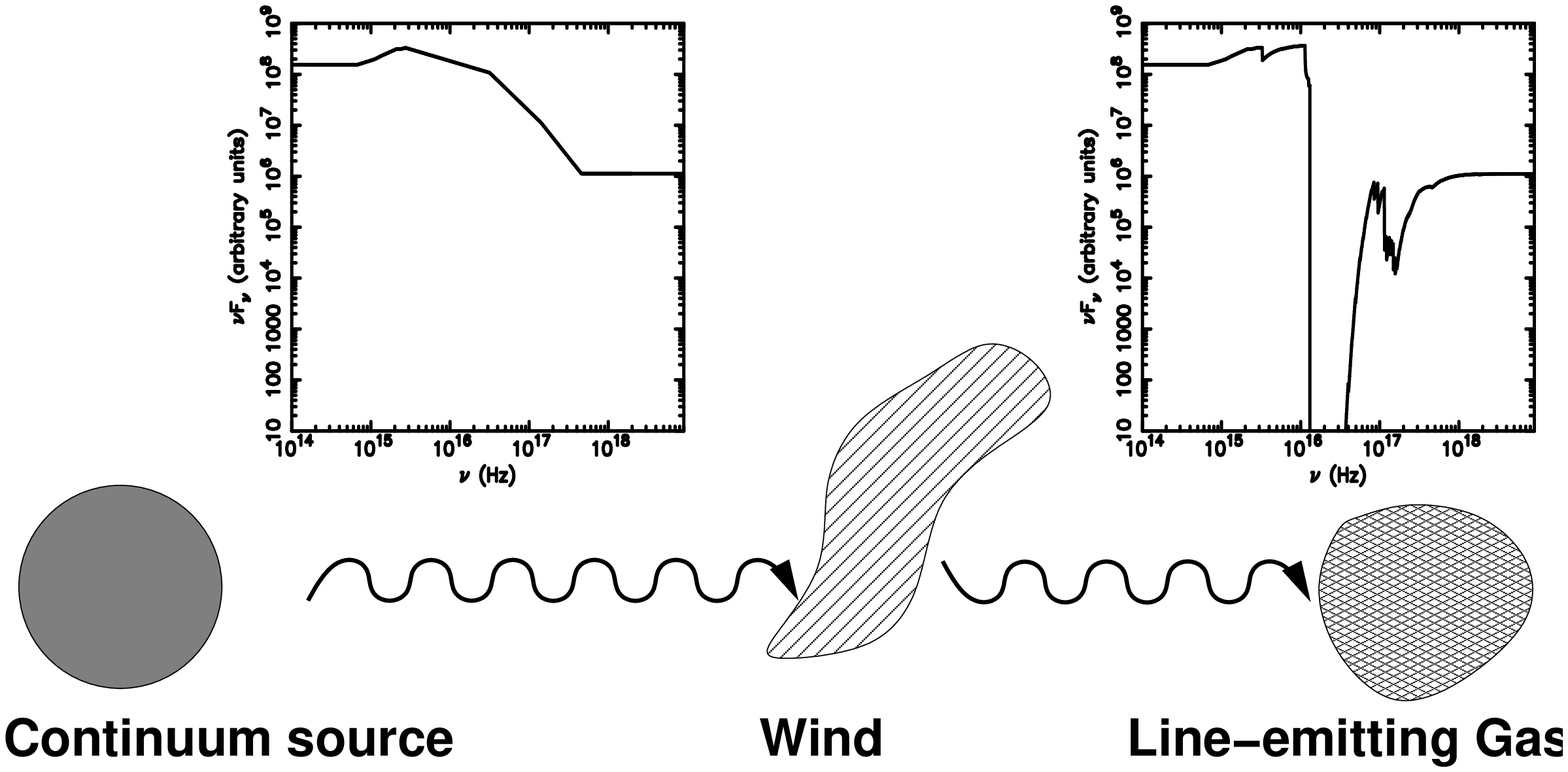}
\caption{\small 
{A schematic illustration of the concept of ``filtering''.  The
  spectral energy distribution originates at the continuum source
  (left) and transverses the wind (center) that responds by emitting
  blueshifted lines.  The transmitted continuum then
  illuminates gas   emitting the intermediate- and low-ionization
  lines (right).   Because the   helium continuum is missing, the
  characteristic ionization of the lines is lower than it would have
  been without   filtering.}} 
\label{fig2}
\end{center}
\end{figure}

We observed two NSL1s, IRAS~13224$-$3809 and 1H~0707$-$495, using {\it
 HST} in 1999 \citep{lm04}. The spectra are characterized by very blue
 continua;   broad, strongly blueshifted high-ionization lines
 (including  \ion{C}{IV} and \ion{N}{V}); and narrow, symmetric
 intermediate-  (including \ion{C}{III}],  \ion{Si}{III}], and
 \ion{Al}{III}), and  low-ionization (e.g.,  \ion{Mg}{II})  lines
 centered at the rest  wavelength.  As discussed in \S 2,
 high-ionization lines are commonly  blueshifted with respect to
 intermediate- and low-ionization lines,  but these two objects
 exhibited an extreme example of this  phenomenon in which the lines
 are almost completely decoupled.   The  emission-line profiles
 suggest that the high-ionization lines are  produced in a wind and
 that the intermediate- and low-ionization  lines are produced in
 low-velocity gas associated with the accretion  disk or base of the
 wind.  The highly asymmetric profile of  \ion{C}{IV} suggested that
 it is dominated by emission from the wind,  so we developed a
 template from the \ion{C}{IV} line. We modeled   the bright emission lines
 in the spectra using a combination of this  template and a narrow
 symmetric line centered at the rest  wavelength \citep{lm04}.  

Next, photoionization modeling using {\it Cloudy} was performed for
the  broad blueshifted wind lines and the narrow, symmetric,
rest-wavelength-centered disk lines separately \citep[for details,
  see][]{leighly04}.  A broad range of physical conditions was
explored for the wind component, and a figure of merit was used to
quantitatively evaluate the simulation results.   The wind lines were
characterized by relatively strong \ion{N}{V} and 1400\AA\/ feature,
but weak \ion{C}{IV} feature.  At first glance, this was difficult to explain
because \ion{C}{IV} falls between the other two in terms of ionization
potential. A somewhat X-ray weak continuum and elevated metallicity
produced an acceptable model for a broad range of densities $\log n
\sim 7-11$) and ionization parameters ($\log U \sim -1.2$ to $-0.2$)
and a small column density of $\log N_H \sim 21.4$. These parameters
work because the lower X-ray flux reduces heating, and 
the higher metallicity allows the gas to cool more efficiently,
allowing \ion{C}{IV} and \ion{O}{IV}] (a component of the 1400\AA\/
feature) to remain strong in the region of parameter space where
\ion{N}{V} is strong.  

We then constructed a photoionization model for the intermediate- and
low-ionization ``disk'' lines. The disk lines include \ion{C}{III}]
 and \ion{Si}{III}], so the density could be constrained. The \ion{C}{IV}
line had little kinematic overlap with the intermediate- and
low-ionization lines.  This seemed to imply that the
ionization state of the gas was rather low ($\log U \sim -3$), so that
it would not produce narrow \ion{C}{IV}), which in turn placed the
emitting gas at a very large distance from the central engine that
seemed incompatible with the width of the lines.  However, we realized that
the radius of the disk-line-emitting gas would be smaller 
if the continuum were first transmitted through the wind before it
illuminated the gas emitting those lines.  The concept is illustrated
in Fig.\ 2.    Transmission through the wind removes the photons in
the helium continuum, and therefore naturally reduces the \ion{C}{IV}
and other high-ionization line emission from the disk component.  We
found that we could produce the observed lines with a much larger
ionization parameter of $\log U\sim -2.1$, and thus with a radius
smaller by a factor of $\sim 2.8$, and a velocity width larger by a
factor of $\sim 1.7$.  

As discussed by \citet{leighly04} the concept of filtering may be
quite general.  For example, filtering may explain why objects with
blueshifted high-ionization lines have strong low-ionization lines such
as \ion{Si}{II} \citep{wills99}.  It may explain the stronger
\ion{Fe}{II} emission in BALQSOs \citep{weymann91}.  In addition,
since the characteristic line-emission radius is reduced, it could
skew black-hole mass estimates. 


\acknowledgements KML and DC thank the organizers for the
opportunity to present this review.  Part of the work in writing this
talk and review was done while KML was on sabbatical at the 
Department of Astronomy at The Ohio State University, and she thanks
the members of the department for their hospitality.  KML and DC
acknowledge NNG05GB38G, NNG05GD73G and NNG05GD01G for support.


\end{document}